\documentclass[final,5p,times]{elsarticle}

\usepackage{lineno,hyperref}
\usepackage{comment}
\modulolinenumbers[5]

\journal{Physics Letters A}

\usepackage{amssymb}
\usepackage{amsmath}
\usepackage{amsfonts}
\usepackage{physics}
\usepackage{comment}

\begin{document}

\begin{frontmatter}




\title{Coherence resonance control via nonlocal coupling in an ensemble of non-excitable oscillators}


\author{Aleksey Ryabov}

\author{Vladimir V. Semenov}
\ead{semenov.v.v.ssu@gmail.com}
\address{Institute of Physics, Saratov State University, 83 Astrakhanskaya Street, Saratov, 410012, Russia}

\date{\today}

\begin{abstract}
Nonlocal interaction is shown to be an appropriate tool for controlling coherence resonance in ensembles of non-excitable oscillators. The constructive role of nonlocal coupling is demonstrated  through numerical simulations on an example of coupled generalized Van der Pol oscillators close to the saddle-node bifurcation of limit cycles. In particular, increasing the coupling radius is found to enhance coherence resonance, which manifests itself in the evolution of power spectra and dependencies of the correlation time on the noise intensity. Nonlocal coupling is interpreted as an intermediate topology between local and global coupling, both of which are also examined as mechanisms for controlling coherence resonance.
\end{abstract}




\end{frontmatter}

\section{Introduction}
\label{intro}
Coherence resonance represents an interdisciplinary phenomenon exhibited by excitable \cite{pikovsky1997,lindner1999,lindner2004,deville2005,muratov2005,semenov2017,semenov2018} and non-excitable  \cite{gang1993,ushakov2005,zakharova2010,zakharova2013,geffert2014,semenov2015} systems.  Stochastic processes associated with coherence resonance are observed across various fields, including neurodynamics \cite{pikovsky1997,lee1998,lindner2004,pisarchik2023,tateno2004}, microwave \cite{dmitriev2011} and semiconductor \cite{hizanidis2006,huang2014,shao2018} electronics, optics \cite{dubbeldam1999,giacomelli2000,avila2004,otto2014,arteaga2007,arecchi2009}, quantum physics \cite{kato2021}, thermoacoustics \cite{kabiraj2015}, plasma physics \cite{shaw2015}, hydrodynamics \cite{zhu2019}, climatology \cite{bosio2023} and chemistry \cite{miyakawa2002,beato2008,simakov2013}. Besides manifesting on its own, coherence resonance can co-occur with another phenomena, such as synchronization \cite{balanov2009,semenov2025-2} and noise-induced transitions \cite{semenov2017}. Furthermore, coherence resonance oscillators can exhibit mutual synchronication or be synchronized by external periodic forcing \cite{han1999,ciszak2003,ciszak2004,korneev2025}. Notably, the synchronization of the noise-induced oscillations occurs similarly as compared to deterministic quasiperiodic systems \cite{astakhov2011}.

The broad range of dynamical systems exhibiting coherence resonance and related applications motivates search for coherence resonance control schemes. Various approaches have been proposed and successfully applied to regulate the degree of the coherence resonance manifestation or to shift optimal values of noise intensity to a more suitable range. In particular, the use of time-delayed feedback enables controlling the properties of noise-induced oscillations in systems with type-I \cite{aust2010} and type-II \cite{janson2004,brandstetter2010} excitability as well as in non-excitable \cite{geffert2014,semenov2015} coherence resonance oscillators. Furthermore, the noise's characteristics can be tuned to enhance or to suppress coherence resonance. For instance, this can be achieved through adjusting the correlation time of coloured noise \cite{brandstetter2010}. Coherence resonance can also be effeciently controlled by varying the parameters of Lévy noise (the noise's probability density function transforms when parameters change) which was demonstrated both numerically and experimentally \cite{semenov2024}. 

In networks of coupled oscillators, coherence resonance can be controlled by adjusting the coupling strength and the network topology, as shown, for instance, in multilayer multiplex networks \cite{semenova2018,masoliver2021} and single-layer networks of excitable oscillators with nonlocal interactions \cite{ryabov2025,bukh2025}. In the current paper, the issue of nonlocal-coupling-based control of coherence resonance is addressed on an example of networked non-excitable oscillators. Comparative analysis of results obtained for networks of excitable (see Refs. \cite{ryabov2025,bukh2025}) and non-excitable (considered in the present paper) oscillators will provide insight into how the intrinsic peculiarities of the network element behaviour affect the collective dynamics coherence as the coupling strength and coupling radius vary.


\section{Model and methods}
\label{model_and_methods}
We investigate coupled identical generalized Van der Pol oscillators with fifth-order nonlinearity. The ensemble model under study is defined by the following equations:
\begin{equation}
\label{eq:general}
\begin{array}{l}
\dfrac{d^2 x_{i}}{dt^2}-\left[\varepsilon +\mu x_i^2-x_i^4 \right]\dfrac{dx_i}{dt}+\omega_0^2x_i \\
\\
=\sqrt{2D}n_i(t)+f_i(x_1,x_2,...,x_N),
\end{array}
\end{equation}
where $i=1,2,...,N$ ($N=100$ is the total number of interacting oscillators), $x_i$ are the dimensionless dynamical variables, $t$ is the dimensionless time, parameters $\varepsilon \in \mathbb{R}$ and $\mu>0$ are  responsible for the oscillators' excitation and dissipation, respectively, $\omega_0=1$ is the oscillators' eigenfrequency near the Andronov-Hopf bifurcation, noise terms $\sqrt{2D}n_i(t)\in\mathbb{R}$ refer to white Gaussian noise of intensity $D$. The noise sources are assumed to be statistically independent, i.e., $<n_i(t)>=0$ and $<n_i(t)n_{j}(t)>=\delta_{ij}\delta(t-t')$, $\forall i,j$. Terms $f_i(x_1,x_2,...,x_N)$ are responsible for the action of coupling. Two options for the coupling terms are studied:
\begin{equation}
\label{eq:nonlocal_coupling}
\begin{array}{l}
f_i(x_1,x_2,...,x_N)= \dfrac{\sigma}{2R} \sum\limits_{j=i-R}^{i+R}(x_j-x_i),
\end{array}
\end{equation}
\begin{equation}
\label{eq:global_coupling}
\begin{array}{l}
f_i(x_1,x_2,...,x_N)= \dfrac{\sigma}{N} \sum\limits_{j=1}^{N}(x_j-x_i),
\end{array}
\end{equation}
where the coupling strength and radius are denoted by $\sigma$ and $R$, respectively. Terms (\ref{eq:nonlocal_coupling}) describe the local interaction at $R=1$ and the nonlocal one at $R>1$, whereas terms (\ref{eq:global_coupling}) correspond to global coupling.

\begin{figure}[t]
\centering
\includegraphics[width=0.35\textwidth]{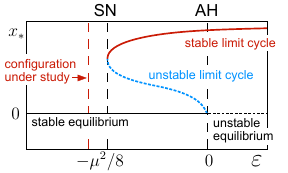}
\caption{Schematically illustrated bifurcation diagram of the coupling- and noise-free oscillators (see Eqs.(\ref{eq:general}) at $D=0$ and $f_i(x_1,x_2,...,x_N)\equiv 0$). The black dashed lines SN and AH indicate the saddle-node and subcritical Andronov-Hopf bifurcation, correspondingly. The red dashed line corresponds to the parameter set considered in the current paper ($\varepsilon=-0.04$, $\mu=0.5$). For more detailed description of the bifurcation analysis results, see Ref. \cite{semenov2015}.}
\label{fig1}
\end{figure}

Due to the presence of nonlinear friction described by the fourth-order function, the coupling-free oscillators (i.e. in the absence of terms $f_i(x_1,x_2,...,x_N)$ in Eqs. (\ref{eq:general})) exhibit hard self-oscillation excitation such that two limit cycles, stable and unstable, coexist in the phase space at appropriate parameter values. When varying parameter $\varepsilon$, the regime of coexistence of two limit cycles is realized in the range bounded by a saddle-node bifurcation of limit cycles at $\varepsilon=-\mu^2/8$ and by a subcritical Andronov-Hopf bifurcation at $\varepsilon=0$ (see Fig.~\ref{fig1}). In the range $-\mu^2/8<\varepsilon<0$ the generalized Van der Pol oscillator is bistable, since the second attractor, a stable focus in the origin, coexists with a stable limit cycle. The stable focus is the only attractor in the phase space at $\varepsilon<-\mu^2/8$. It becomes unstable for $\varepsilon>0$ in a subcritical Andronov-Hopf bifurcation at $\varepsilon=0$. In the following, coupled oscillators (\ref{eq:general}) are considered in the monostable steady-state regime at $\varepsilon<-\mu^2/8$ (see the red dashed line in Fig. \ref{fig1}), in particular, at $\varepsilon=-0.04$ and $\mu=0.5$. In such a case, the single generalized Van der Pol oscillator exhibits coherence resonance when increasing the noise intensity \cite{zakharova2010,semenov2015}.

Ensemble (\ref{eq:general}) was numerically modelled by integrating the ordinary differential equations using the Heun method \cite{mannella2002}. The integration method parameters are the time step $\Delta t=0.001$ and the total integration time $t_{\text{total}}=2\times 10^4$. Random initial conditions were used: $x_i(t=0)$ and $\frac{dx_i}{dt} (t=0)$ are random quantities uniformly distributed in the range from -1 to 1. To characterise the properties of coherence resonance, we consider how the stochastic dynamics (time series $x_i(t)$) evolves in response to increasing noise intensity and varying coupling parameters. This implies calculation of the correlation time, $t^{\text{cor}}_i$, characterising the dynamics of each oscillator:
\begin{equation}
\label{eq:t_cor_and_R_ISI} 
\begin{array}{l}
t^{\text{cor}}_i=\dfrac{1}{\Psi(0)}\int\limits_{0}^{\infty} \left| \Psi(s) \right|ds, 
\end{array}
\end{equation} 
where $\Psi(s)$ and $\Psi(0)$ are the autocorrelation function and the variance of the time realization $x_i(t)$. Finally, the averaged value of the introduced quantity is extracted to globally describe the stochastic collective behaviour: $\overline{t}_{\text{cor}}=\dfrac{1}{N}\sum\limits_{i=1}^{N}t^{\text{cor}}_i$. Similarly, the power spectrum averaged over the ensemble, $\overline{S}(\omega)$, is considered to highlight the similar character of the coherence resonance occurrence in single and networked oscillators: $\overline{S}(\omega)=\dfrac{1}{N}\sum\limits_{i=1}^N S_i(\omega)$, where $S_i(\omega)$ is the power spectrum of the network node oscillations $x_i(t)$.

\section{Local and global coupling}
We start by considering a ring of locally coupled oscillators (see model (\ref{eq:general}), where the coupling is specified by Eqs. (\ref{eq:nonlocal_coupling}) with $R=1$). Expectedly, the weak coupling has no significant effect on the collective stochastic dynamics: the interacting oscillator behaviour is largely independent and asynchronous (see the space-time plots in Fig. \ref{fig1}~(a)). However, increasing the coupling strength does not impact the oscillation spatial coherence: if the coupling strength is large and the noise intensity is chosen such that the single oscillators (see Eqs. (\ref{eq:general}) in the absence of the coupling terms) exhibit coherence resonance, the stochastic oscillations remain to be asynchronous  [Fig. \ref{fig1}~(b)]. However, the coupling strength growth causes another effect which consists in the suppression of coherence resonance. In particular, one observes decreasing the local dynamics regularity when increasing the coupling strength. This effect is illustrated in Fig.~\ref{fig1}~(c) on an example of dependencies of the averaged correlation time on the noise intensity obtained at various coupling strengths. Indeed, as shown in Fig.~\ref{fig1}~(c), growth of the coupling strength allows to significantly decrease the peak value of $\bar{t}_{\text{cor}}$ as compared to the dynamics of the single oscillators. At the same time, there are no significant changes in the optimal noise intensity $D_{\text{opt}}$ corresponding to the most coherent oscillations. Nevertheless, the transformation of curves $\bar{t}_{\text{cor}}(D)$ illustrated in Fig.~\ref{fig1}~(c) indicate the option for coherence resonance control at fixed noise intensity. For instance, increasing the coupling strength at $D=0.002$ (the vertical dashed line in Fig.~\ref{fig1}~(c)) suppresses coherence resonance. The described character of the local coupling influence fully correlates with the transformations of the power spectra obtained at fixed noise intensity and increasing $\sigma$ (see Fig.~\ref{fig1}~(d) corresponding to $D=0.002$ and $\sigma \in [0:1]$). It is clearly visible that the decreasing values of $\bar{t}_{\text{cor}}$ correspond to the less prominent main spectral peak (the height of the main spectral peak decreases, while the peak itself broadens and even splits into two ones). Thus, the evolution of quantity $\bar{t}_{\text{cor}}$ is in complete accordance with the transformations of power spectra. 

\begin{figure}[t]
\centering
\includegraphics[width=0.48\textwidth]{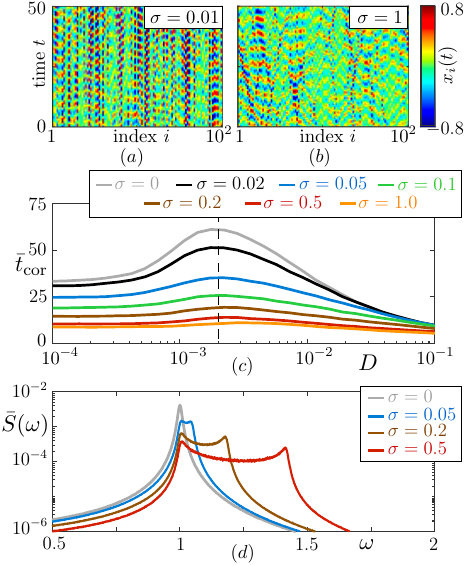}
\caption{Suppression of coherence resonance in ensemble (\ref{eq:general}) due to the action of local coupling defined by Eqs. (\ref{eq:nonlocal_coupling}) at $R=1$: (a)-(b) Space-time plots illustrating the stochastic dynamics as the coupling strength increases at fixed $D=0.002$; (c) Evolution of dependencies of the averaged correlation time on the noise intensity, as the coupling strength increases; (d) Transformation of the averaged power spectrum of oscillations $x_i(t)$ at fixed noise intensity, $D=0.002$ (corresponds to the vertical dashed line in panel (c)), and increasing coupling strength. The oscillators' parameters are $\varepsilon=-0.04$, $\mu=0.5$, $\omega_0=1$.}
\label{fig2}
\end{figure}

\begin{figure}[t]
\centering
\includegraphics[width=0.48\textwidth]{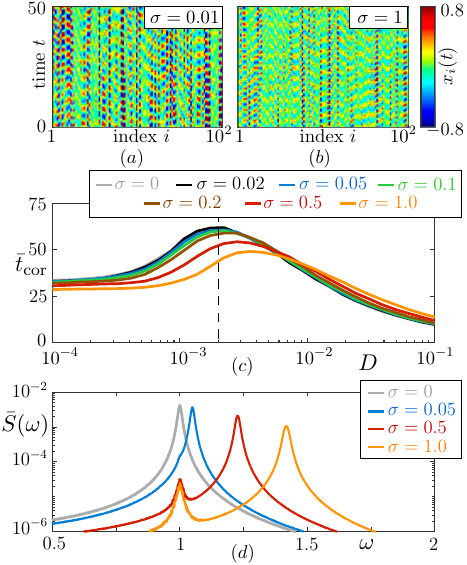}
\caption{Global-coupling-induced suppression of coherence resonance (see ensemble (\ref{eq:general}) with coupling terms (\ref{eq:global_coupling})): (a)-(b) Space-time plots illustrating the stochastic dynamics as the coupling strength increases at fixed $D=0.002$; (c) Evolution of dependencies of the averaged correlation time on the noise intensity, as the coupling strength increases; (d) Transformation of the averaged power spectrum of oscillations $x_i(t)$ at fixed noise intensity, $D=0.002$ (corresponds to the vertical dashed line in panel (c)), and increasing coupling strength. The oscillators' parameters are $\varepsilon=-0.04$, $\mu=0.5$, $\omega_0=1$.}
\label{fig3}
\end{figure}

Similarly to the influence of local interaction, global coupling also provides for suppression of coherence resonance, which is not accompanied by the effects of the noise-induced oscillation synchronization [Fig. \ref{fig3}~(a),(b)]. As illustrated in Fig. \ref{fig3}~(c) for ensemble (\ref{eq:general}) with coupling terms defined by Eqs. (\ref{eq:global_coupling}), the presence of global coupling enables to decrease the peak values of the correlation time. As the strength of global coupling increases, the main spectral peak shifts toward higher frequencies, while its height decreases [Fig. \ref{fig3}~(d)]. It is important to note that the second spectral peak of low amplitude appears at $\omega=1$ when the coupling strength is large enough.
One can notice that a fundamental distinction between the effects of local and global coupling emerges at high coupling strengths. As the strength of global coupling increases, suppression of coherence resonance occurs slower as compared to the system with local interactions (compare Fig.~\ref{fig2}~(c) and Fig.~\ref{fig3}~(c)). As demonstrated in the next section, this established aspect can play a key role and results in coherence resonance control when increasing the radius of nonlocal coupling.

\section{Nonlocal coupling}
\label{sec_nonlocal}
The comparative analysis of dependencies in Fig.~\ref{fig2}~(c) and Fig.~\ref{fig3}~(c) allows to conclude that, at the same coupling strength, the suppression of coherence resonance is less pronounced in the presence of global coupling as compared to the system with local interactions. In particular, considering the curves in Fig.~\ref{fig2}~(c) and Fig.~\ref{fig3}~(c) corresponding to the same coupling strength, one can establish that the action of global coupling gives rise to higher peak values of $\bar{t}_{\text{cor}}$. Based on this fact, it is reasonable to anticipate that increasing the radius of nonlocal coupling (is considered as a continuous transition from local to global coupling) could lead to enhancement of coherence resonance. To test this hypothesis, we further investigate ensemble (\ref{eq:general}) with coupling terms (\ref{eq:nonlocal_coupling}), varying $R$ while keeping $\sigma$ fixed.

\begin{figure}[t!]
\centering
\includegraphics[width=0.49\textwidth]{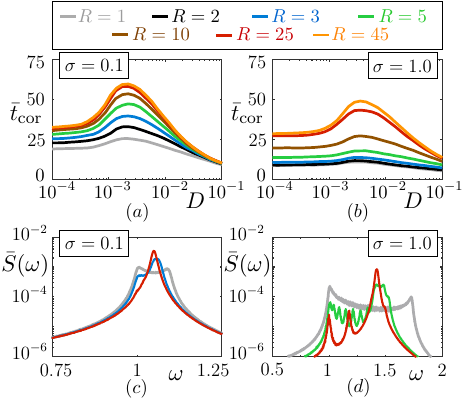}
\caption{Enhancement of coherence resonance when increasing the coupling radius in model (\ref{eq:general}) with nonlocal coupling (\ref{eq:nonlocal_coupling}). The evolution of the dependencies of the averaged correlation time on the noise intensity (panels (a) and (b)) and the power spectrum transformation at $D=0.002$ (panels (c) and (d)) are used to visualise the coherence resonance control at $\sigma=0.1$ (panels (a) and (c)) and $\sigma=1$ (panels (b) and (d)). The oscillators' parameters are $\varepsilon=-0.04$, $\mu=0.5$, $\omega_0=1$.}
\label{fig4}
\end{figure}

The evolution of stochastic dynamics with increasing coupling radius is demonstrated in Fig.~\ref{fig4} for two cases: $\sigma = 0.1$ (panels (a) and (c)) and $\sigma = 1$ (panels (b) and (d)). In both instances, a larger coupling radius leads to enhancement of coherence resonance. The enhancement is manifested as increasing the peak values of $\bar{t}_{\text{cor}}$ corresponding to the most regular oscillations. In the context of power spectra, the coherence resonance enhancement is exhibited as narrowing and rising the main spectral peak at fixed noise intensity and increasing coupling radius [Fig. \ref{fig4} (c),(d)]. The only difference between the cases of low and high coupling strengths consists in the fact that the coherence resonance enhancement has a saturable character in the presence of weak coupling: no noticeable changes in the dynamics are observed when the coupling radius exceeds the value $R = 25$ (see Fig.~\ref{fig4}(a)).

\section{Conclusion}
Control of coherence resonance via nonlocal coupling depends on the intrinsic properties of  the ensemble element dynamics and can be therefore realized in different ways. As reported in Ref. \cite{ryabov2025}, the presence of coupling (local, nonlocal and global) enables enhancing  coherence resonance in ensembles of excitable oscillators as compared to the single oscillator dynamics. At the same time, the efficiency of coherence resonance enhancement depends on the coupling strength, with either local or global coupling topology being more favorable under different conditions. Consequently, varying the coupling radius in ensembles of nonlocally coupled excitable oscillators can lead to either enhancement or suppression of coherence resonance. Notably, a similar effect of nonlocal coupling on stochastic resonance was reported in Ref.~\cite{semenov2025-3}.

In contrast, coupling generalized Van der Pol oscillators (non-excitable systems) results in suppression of coherence resonance in comparison with the behaviour of a single oscillator. However, in the case of global coupling, coherence resonance suppression with increasing the coupling strength is less pronounced. As a result, increasing the coupling radius in ensembles of nonlocally coupled non-excitable oscillators leads to enhancement of coherence resonance. The enhancement of noise-induced coherence is reflected in an increase of the correlation time peak values, which is in full agreement with the corresponding evolution of the power spectra. Thus, similarly to ensembles of excitable oscillators, nonlocal coupling in non-excitable oscillator ensembles represents an effective mechanism for improving the regularity of noise-induced oscillations in the coherence resonance regime.

The presented results are the basis for further studies such as analysis of the appearance of the additional spectral peaks as the strength of local, nonlocal and global coupling increases. Furthermore, it has been demonstrated on examples of globally-coupled oscillators that varying the system size can impact stochastic \cite{pikovsky2002} and coherence \cite{toral2003} resonances. Thus, similar effects are expected to occur with an increase in the coupling radius in ensembles of nonlocally coupled oscillators of different size. However, this problem requires a separate consideration.

\section*{Declaration of Competing Interest}
The authors declare that they have no known competing financial interests or personal relationships that could have appeared to influence the work reported in this paper.

\section*{Acknowledgments}
This work was supported by the Russian Science Foundation (project No. 24-72-00054).

\text{https://rscf.ru/project/24-72-00054/}


\end{document}